\documentclass[graybox,natbib,nosecnum]{svmult}
\bibpunct{(}{)}{;}{a}{}{,} % suppress commas between author-names and year

\usepackage{natbib}
\usepackage{helvet}     % selects Helvetica as sans-serif font
\usepackage{courier}    % selects Courier as typewriter font
\usepackage{type1cm}    % activate if the above 3 fonts are
\usepackage{makeidx}     % allows index generation
\usepackage{graphicx}    % standard LaTeX graphics tool
\usepackage{multicol}    % used for the two-column index
\usepackage[bottom]{footmisc}% places footnotes at page bottom
\usepackage[normalem]{ulem}	% for strike-through of text with \sout{} 
\usepackage{hyperref} %for hyperlinks
\usepackage{longtable}
\usepackage{soul}  % for high-lighting of text
\makeindex       % used for the subject index
\begin{document}

\title*{ Variability of Brown Dwarfs}
\author{\'Etienne Artigau}
\institute{\'Etienne Artigau \at Institut de Recherche sur les Exoplan\`etes (IREx), D\'epartement de Physique, Universit\'e de Montr\'eal, C.P. 6128, Succ. Centre-Ville, Montr\'eal, QC, H3C 3J7, Canada \email{artigau@astro.umontreal.ca} }
\maketitle

\abstract{Brown dwarfs constitute a missing link between low-mass stars and giant planets. Their atmospheres display chemical species typical of planets, and one could wonder whether they also have weather-like patterns. While brown dwarf surface features cannot be directly resolved, the photometric and spectroscopic modulations induced by these features, as they rotate in and out of view, provide a wealth of information on the evolution of their atmosphere. A review of brown dwarfs variability through the L, T and Y spectral types sequence is presented, as well as the constraints that they set on the nature of weather-like patterns on their surface.}

\section{Introduction: The Surface Features of Brown Dwarfs}
When a {brown dwarf} (BD) is represented for public outreach purpose, it is most often shown as displaying large-scale atmospheric features such as bands and storms, not unlike those of the Solar System  giants. While such a representation may seen like an educated guess for an object bearing many of the hallmark molecules found in planets (e.g., water, methane, carbon monoxide and, in the coolest brown dwarfs, ammonia), the true appearance of BDs remains largely unknown. While this question may appear trivial at a first glance, it is a key element in accurately describing these objects. Models attempting to fit the spectral energy distribution of BDs assume a uniform set of physical parameters (surface gravity, temperature, dust settling, chemical composition, etc). However, this emerging flux includes contributions from various regions with a range of physical properties. The 5\,$\mu$m flux distribution on the surface of Jupiter provides a nearby example of the limitation of disk-averaged {photometry}; a large fraction of the flux arises from hot spots that cover only a small fraction of the surface.
 
As for most stellar objects, BDs cannot be resolved with existing nor any upcoming facilities. A few of the very nearest {M dwarfs} have been resolved through interferometry (e.g., \citealt{Boyajian2012}) and the technique may eventually be extended to cooler objects, but this falls short of providing a surface map. Interferometry is mostly used to provide accurate angular diameters that, combined with a distance measurement, yield a true physical size. The question regarding the presence of large-scale storms, bands or weather-like features on BDs thus will not be answered directly in the foreseeable future. 

\begin{figure}
\includegraphics[width=\textwidth]{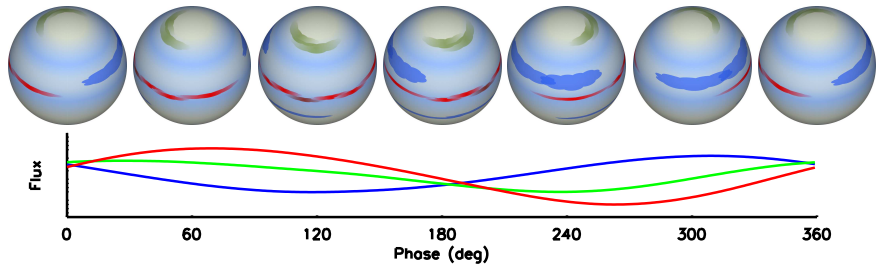}
\caption{Model view of rotation-induced modulation at different wavelengths. The toy brown dwarf displays  weather patterns on various spatial scales and colors. As these patterns rotate in and out of view, the lightcurve  displays a wavelength-dependent modulation. Phase lags between wavelengths and varying amplitude ratio hint at large-scale color differences.}
\label{demo}    % Give a unique label
\end{figure}

It was suggested shortly after the discovery of the first BD {\citep{Nakajima1995, Rebolo:1995}} that weather-like patterns on their surface may lead to rotation-induced variability (e.g., \citealt{Tinney1999}), providing a first glance at the diversity of surface features in these objects. The non-detection of variability does not necessarily disprove the existence of weather-like features on BDs as these can be much smaller than the diameter of the object, or distributed in bands that have little longitudinal structure, but a variability detection would set constraints on surface  inhomogeneities.  While BDs may display complex cloud patterns, as show in Figure~\ref{demo}, rotation-induced variability only probes the largest structures and informs the observer on how these structures contrast against the rotation-averaged flux. For comparison, Solar System giants  observed as unresolved point sources  display rotation-induced optical variability of 1-2\% to $\sim$4\% in the case of Neptune and Jupiter, respectively \citep{Simon:2016}. The mid-infrared variability of Jupiter is much larger ($\sim20$\,\%) as the flux is largely emitted from hot spots irregularly distributed on its surface \citep{Gelino:2000}. 

A comparison of BDs with Solar System giants has a limited validity and should always be done with the appropriate caveats. Heat transport drives weather in the Solar System giants, but the heat transported per unit surface on a BD is $10^3$ (late T dwarf) to $10^5$ (early L dwarf) times larger than that of Jupiter. Furthermore, stellar-like activity extends at least to mid L dwarfs (e.g., \citealt{Mohanty:2003}), hence spot-induced modulation may be expected for these brown dwarfs.  The atmospheres of {T dwarfs} are cool enough that their ionization levels are very low and they are thus expected to be decoupled from the BD's magnetic fields, but objects as cool as T6 have nevertheless shown H$\alpha$ emission of yet uncertain origin \citep{Burgasser2002b}. Surface features on BDs may therefore differ  in nature from those of Solar System giants and may conceivably include a complex interplay between stellar-like activity and planet-like weather patterns.

As one attempts to detect rotation-induced variability, it is essential to have an order-of-magnitude estimate of the timescales involved. Evolutionary models of BDs indicate that their radius is only a weak function of mass and age (e.g., \citealt{Chabrier:2000a}), with all BDs older {than 120\,Myr} having radii between 0.85 and 1.2\,$R_{\rm Jup}$. The {rotational broadening} of BDs can be constrained through high-resolution spectroscopy, and typical $v \sin i$ values range from 10 to 60 km/s among {L dwarfs} \citep{Mohanty:2003}, with a majority of objects rotating faster than 20\,km/s. These values correspond to rotation periods ranging from 2 to 12\,h. These relatively short rotation periods, when compared to most main-sequence stars, imply that one or more rotations can be probed during a single observing night in most cases. This simplifies the observing strategies and alleviates the problems of sparse phase sampling associated with ground-based monitoring on longer timescales.

%Angular size
%Solar system planets' surface diversity 
%\citep{Gelino:2000}
%Na\"ive assumption of single-phased disk
%\section{$J$-band brightnening}
%\citep{Marley2003}

\section{Early Efforts in Brown Dwarf Variability}
The first effort to detect BD variability was reported by \citet{Tinney1999} on two M and L dwarfs. The observations were carried out with a tunable filter probing a strong TiO absorption and a nearby featureless region of the far-red spectrum. The authors reported some evidence of variability for the M9 dwarf LP~944--20, with a gradual increase in flux through the 1.5\,h observing sequence. While setting little additional constraints on the timescale or eventual periodicity of the signal, these observations demonstrated that percent-level differential photometry is possible on these faint objects, paving the way for future work. Concurrent efforts by \citet{Bailer1999} in monitoring 6 late-M and early-L dwarfs in $I$ band led to the detection of periodic 4\% peak-to-peak modulation in the L1.5 2MASSW\,J1145572+231730 at a period of $\sim7$\,h. The observations were performed sparsely over $\sim10$ rotation periods, leading to doubts regarding the true periodicity, if any. This ambiguity was confirmed by \cite{Bailer2001} in a reanalysis of the dataset, in combination with new data for this target, where the apparent periodicity was instead attributed to a sampling effect. Despite the challenges brought by the sparse sampling in these datasets, the authors demonstrated that roughly 50\% of late-M and early-Ls vary at the few percent level in the far-red, but ascribing this variability to a physical cause remained tentative due to the absence of multi-wavelength detections of variability or to a clear trend between target properties and the variability level. These results were overall consistent with those of \citet{Gelino2002} who monitored 18 {L dwarfs} over long time baselines. One notable discovery by \citet{Gelino2002} is the detection of a dimming by $\sim$2\% over 10 days of the L1 dwarf 2MASS\,J$13301009+1411132$. This is too long to be attributed to rotation-induced modulation, and it thus provided a first hint of a possible large and long-lived  storm-like feature on a BD.  The discovery of a 1.8\,h photometric modulation for the L dwarf Kelu-1 \citep{Clarke2002b} and subsequent spectroscopic discovery of H$\alpha$ emission modulation at the same period \citep{Clarke:2003}, suggesting an activity-induced variability.  \citealt{Liu:2005} resolved Kelu-1 as tight binary, complicating the interpretation of unresolved measurements. 

Early effort concentrated on late-M to mid-Ls and did not include later-type objects because no relatively bright T dwarfs were known. This has changed due in large part to the SDSS and 2MASS surveys (e.g.,\citealt{Burgasser1999,Leggett:2000}), allowing and variability searches to be extended to cooler objects. \citet{Enoch2003} performed a $K_s$ survey of L and T dwarfs with a  sparse sampling of a few visits distributed over about a month. Three out of 9 objects were reported as variables with amplitudes ranging from 10\% to 48\%. In retrospect, these large amplitudes should be taken with caution as subsequent surveys found $>4$\% peak-to-peak variability amplitudes to be relatively rare and only one T dwarf in the $>50$ monitored since then has shown $>15$\% near-infrared variability (2MASS\,J21391365-3529507, hereafter 2M2139; \citealt{Radigan:2012}).

The surveys by \citet{Koen:2004} and \citet{Koen:2005} provided the first near-continuous monitoring of a relatively large sample of objects in $J$, $H$ and $K$ bands. The uninterrupted monitoring were sufficiently long to cover the expected median rotation period of brown dwarfs, and led to a better handling of systematics. While no clear detection was reported, the $<2$\% limits ($<1$\% in $J$) on periodic variability provided a significant improvement over previous limits. These results led to the conclusion that either the near-infrared variability of L and T dwarf was typically small (i.e., at the sub-\% level) or that rotation periods from $v \sin i$ measurements were erroneous, which would imply that BDs were generally much slower rotators than previously thought. 

Space-based platforms provide unrivalled stability for variability studies, and brown dwarfs are no exception.  \citealt{Morales2006} obtained the first time-series of brown dwarfs, targeting three late-Ls with Spitzer, using the IRAC instrument at 4.5 and 8\,$\mu$m. For ground-based observatories, this domain suffers from very high background and time-varying atmospheric water absorption, rendering precise photometric measurements nearly impossible. The 6-8\,h times series displayed percent-level near-periodic $4.5$\,$\mu$m photometric modulation. These observations paved the way for large Spitzer BD variability surveys as mentioned below.

\section{The Persistent High-Amplitude Variability of SIMP0136}

\begin{figure}
\includegraphics[width=\textwidth]{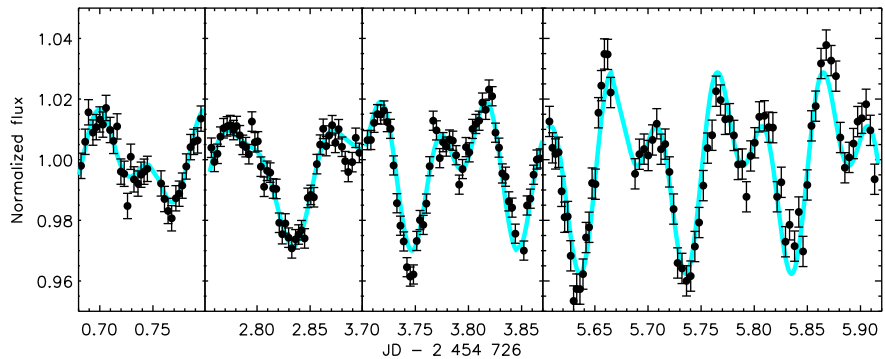}
\caption{
 $J$-band lightcurve of SIMP0136 over four epochs. The cyan curve is two-harmonic sinusoidal fit at a 2.388\,h period for each visit. The lightcurve evolves between epochs, but is well fitted by a periodic signal within each of them. Data from \citet{Artigau2009}.}
\label{fig0136artigau2009}% Give a unique label
\end{figure}

Early claims of variability detection past mid-Ls were not detected at high confidence and thus no object appeared to show consistent and readily detectable variability. This impeded follow-up studies of variability as one needed a reliable variable target in order to plan for detailed follow-up studies. Upon the discovery of the nearby T2.5 SIMP\,J013656.57+093347.3 (SIMP0136; \citealt{Artigau:2006}), it was realized that this object is an ideal target for variability studies. SIMP0136 falls  at the {L/T transition} where dust-bearing clouds form close to the photosphere and large-scale cloud patterns may lead to rotation-induced variability. A set of $J$-band photometric sequences of four nights spread over a week was obtained at the {\it Observatoire du Mont-M\'egantic} 1.6-m telescope \citep{Racine1978}. The photometric sequence displayed an unambiguous near-periodic modulation that evolved from night to night (See figure~\ref{fig0136artigau2009} and \citealt{Artigau2009}). On the two first nights of observation, the photometric modulation was found to be nearly sinusoidal in shape, while it displayed a more complex M-shaped structure at subsequent epochs. Similar structures that were observed on the two last days indicated that weather patterns on this BD can  evolve relatively rapidly and survive for at least dozens of rotation periods.

The variability amplitude of SIMP0136  ($\sim5$\%) and its rapid modulation (2.4\,h) would have been readily detected by observations such as those performed by \citet{Koen:2005, Koen:2004} on a sample that included L/T transition objects. The non-detection of variability in previous surveys implies that this level of variability is uncommon. One noteworthy example is SDSSp\,J$125453.90-012247.4$, which has been observed by various authors \citep{Radigan:2014, Metchev:2015,Wilson:2014a, Koen:2004, Girardin:2013}, with detection thresholds well below 1\%, and has never been reported as variable, despite displaying bulk photometric properties that are very similar to those of SIMP0136. While one could argue that photometric variability could be intermittent, the data in hand suggests otherwise.  \citet{Artigau2009} detected variability of SIMP0136 in September 2008 and follow-up observations in December 2008 showed a similar level of variability. Subsequent observations by various teams \citep{Wilson:2014a, Radigan:2014, Croll:2016, Apai:2013, Metchev:2013} confirmed its variability with a $\sim2.4$\,h modulation, albeit at a varying amplitude. These observations span a total of 8 years and there is no reported non-detection of variability for this object.

\section{High-Accuracy Monitoring of Large Samples of L and T Dwarfs\label{highaccuracy_monitoring}}

One of the important questions regarding the variability of BDs is its relation with the L/T transition. Does the gradual sinking of dust-bearing clouds below the photosphere in this spectral range lead to an increased variability due to holes in clouds (e.g., \citealt{Marley2003})? This question can only be answered through a survey of a large sample of BDs, from early-L to late-T dwarfs. Various teams gathered observations of large BD samples in an attempt to answer this question. The largest samples of high-precision (sub-\%) $J$-band observations were obtained by \citet{Radigan:2014}, \citet{Wilson:2014a} and its reanalysis by \citeauthor{Radigan:2014a} (respectively 62 and 69 BDs), while \citet{Metchev:2015} surveyed a large sample (44 objects)  with {SPITZER} at 3.6 and 4.5\,$\mu$m. The approaches of these surveys are not identical, so a one-to-one comparison should be done with care. The \citet{Metchev:2015} observations were much longer ($\sim$$24$\,h) than the two other surveys ($2-6$\,h). Furthermore,  different wavelength regions were probed, implying that differences in photometric amplitudes cannot be directly compared. 

The \citet{Radigan:2014} survey was mostly performed at the Du Pont 2.5\,m telescope with a filter that purposely avoided the deep telluric water absorption bands longward of $1.35\mu$m, minimizing non-grey absorption effects from the Earth's atmosphere. The survey led to the detection of significant variability in 9 out of 57 L4$-$T9 BDs where all of the high-amplitude variables ($>2$\,\%) are  within the L9$-$T3.5 range. When accounting for projection effects that dilute the signal, the fraction of objects that would be $>2$\,\% variables as seen from an equatorial viewpoint raises to $\sim$80\,\%. These results led to the conclusion that high-amplitude variability is indeed more common at the L/T transition.

\citet{Wilson:2014a} presented a similar dataset obtained at the ESO 3.6\,m NTT telescope, albeit with typically shorter ($2-4$\,h) monitoring. The results suggested that the fraction of variable objects within the L/T transition is similar to that of objects outside the transition, with a nearly uniform distribution of variable objects through the early-L to late-T sequence. This result was called into question by \citet{Radigan:2014a} in which a  reanalysis of the dataset was performed.  Most early and mid-L variability detections were found to be constant at the sub-\% level in this new analysis. A  statistical assessment of combined datasets, with 82 individual objects, indicated a fraction  of high-amplitude variables of $24^{+11}_{-9}$\,\% within the {L/T transition}; much larger than the $\sim$$3$\,\% fraction  of variables outside of it.

The \citet{Metchev:2015} sample does not display a significant increase in  the fraction of high-amplitude variables close to the L/T transition compared to earlier or later objects. The very high level of stability of {Spitzer} allowed the detection of very low level variability in the brighter - and typically  earlier spectral types - objects. These observations showed that nearly two thirds of L dwarfs display $0.2-2\%$ variability and about half of these variables are irregular, showing an evolution of surface features on timescales of a few hours. The steady increase in the maximum amplitude detected  with spectral type at both 3.6 and  4.5\,$\mu$m is also notable. No L dwarf was found to vary by more than $2$\,\%, while some T dwarfs vary by up to  4.6\,\%.  The absence of  a clear relation between the 3.6 or  4.5\,$\mu$m amplitudes is also noteworthy. If a single type of clouds were contrasting against a typical background, one would expect the contrast ratio between the two wavelength regions to be constant.  Among both L and T variables, the  4.5\,$\mu$m  to 3.6\,$\mu$m  amplitude ratio  varies from $<0.5$ to $\sim$$2$, with no discernible trend as a function of color or spectral type.

While most contributions to the field concentrated on the occurrence and color-dependency of variability, the evolution of the light curves also sheds light on the physical distribution of atmospheric features inducing variability. \citealt{Apai:2017} presented a monitoring of three known bright variable T2 dwarfs (2M2139, SIMP0136 and 2M1324). The {Spitzer} monitoring sequences were spread over more than a year, probing light curve evolution over timescales of more than 1000 rotations. The variability of objects was detected at very high significance, and the curve displayed a near-periodic modulation with a changing shape and amplitude. The probability distribution of fitted periods was found to display discrete peaks attributed to individual bands with differing wind velocities, analogous to those found on Neptune. The wind velocities recovered from the light curve inversions ranged from 550 to 800\,m/s for all three BDs, commensurate with wind velocities in those of Neptune (300-400\,m/s).

\section{Surface Gravity and Brown Dwarf Variability}

\begin{figure}
\includegraphics[width=\textwidth]{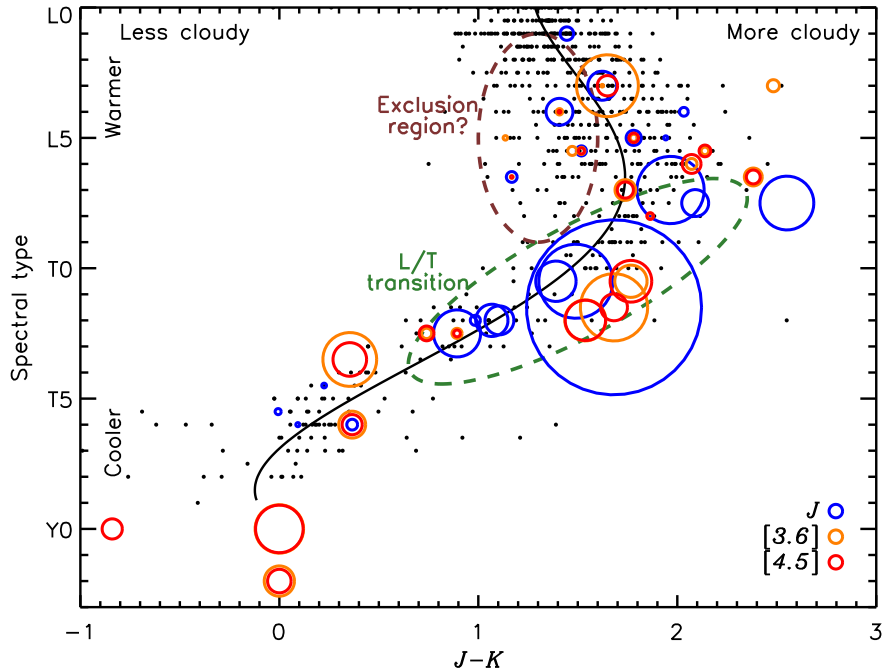}
\caption{ Compilation of variable L, T and {Y dwarfs} in the literature in a $J-K$ versus spectral type diagram (open color circles), with field objects (black dots). The symbol size is proportional to the variability amplitude, the most variable of all being the T1.5 dwarf 2M2139 (26\% in $J$). The largest reported amplitude is shown for objects with more than one detection in a given bandpass. Y dwarfs that do not yet have a published $K$-band magnitude have been arbitrarily set at $J-K=0$ for display purpose. The blue, orange and red circles respectively denote variability detections in $J$, and the $3.6$\,$\mu$m and $4.5$\,$\mu$m Spitzer IRAC bandpasses. A third order polynomial fit to the  color/spectral type relation is also shown. Most high-amplitude variables fall within the L/T transition region, where variability is expected  as cloud decks sink below the photosphere. Interestingly, no high-amplitude variable has been found among L dwarfs bluer than the average for their spectral type. This exclusion region suggests that only redder, and generally lower-gravity, L dwarfs can display $>2$\% variability in the near and mid-IR. Data from \citet{Artigau2009, Radigan:2014,Yang:2015,Metchev:2015,Radigan:2014a,Biller:2015,Buenzli:2015,Clarke:2008, Lew:2016,Buenzli:2015,Cushing:2016,Leggett:2016a,Croll:2016a}. The field brown dwarf photometric data is from \citet{Gagne:2015} and publicly available  at \href{www.astro.umontreal.ca/~gagne/listLTYs.php}{www.astro.umontreal.ca/~gagne/listLTYs.php}. }
\label{compil} 
\end{figure}

Surface gravity in BDs has a direct influence on the behavior of dust. Lower surface gravities in {L dwarfs} lead to a slower dust settling rate, which results in thicker cloud decks at high altitudes and redder near-infrared colors. As dust-bearing clouds play a central role in brown dwarf variability, it is natural to expect a correlation   in the variability level with surface gravity. \citet{Metchev:2015} explicitly included low-gravity {L dwarfs} in their sample to explore this possibility. They found that while the fraction of variable objects is found to be the same within the low-gravity and field sub-samples, the low-gravity objects tend to display higher variability amplitudes. A few variability searches have targeted single objects with peculiar properties, including very-low gravity {L dwarfs}. \citet{Biller:2015} measured a $J$-band variability at the $7-10$\% level for the isolated planetary-mass late-L dwarf PSO\,J$318.5-22$. While this contribution reports on a single object within an ongoing program, it is a remarkable result as the large surveys described earlier failed to uncover L dwarfs that vary at $>4$\%.  The subsequent detection by \citet{Lew:2016} of a $\sim$8\% variability in the planetary-mass L dwarf WISEP\,J$004701.06+680352.1$ (W0047) with WFC3 grism observations seems to further confirm the link between high-amplitude variability and low gravity. Figure~\ref{compil} compiles all $J$, 3.6\,$\mu$m and 4.5\,$\mu$m variability detections in a spectral-type versus color diagram. The detections to date  suggest a higher fraction of high-amplitude variables among very red Ls, but one must bear in mind that some of these discoveries were from surveys explicitly targeting very red low-gravity objects.

While surface gravity may be the link between color and variability amplitude, other physical parameters may explain this correlation. Rotation-induced variability is maximal for inclination close to 90$^{\circ}$; if brown dwarfs have colors that differ at the equator relative to the poles, than the unresolved color of an object will correlate with its inclination, providing an alternative explanation for the correlation described earlier. Through high-resolution spectroscopy, \citealt{Vos:2017a} measured the projected rotational velocity of a sample of early-L to early-T dwarfs with known rotation periods, thus constraining their inclination to the line of sight. The sample showed a significant correlation between the $J-K$ color anomaly (an object's color relative to the mean color of objects of a similar spectral type) and its inclination; redder objects having higher inclination (i.e., seen by the equator). This results implies that on average, brown dwarf equators, at least in this spectral type range, are on average redder than their poles. This is a first hint at the latitudinal variations in brown dwarf properties, complementary to the longitudinal inhomogeneities probed through rotation-induced variability. Furthermore, as expected from geometric arguments, the authors found a correlation between variability amplitude and correlation, which also translates into a correlation between variability amplitude and color. To what extent low surface gravity and viewing angle respectively contribute to the observed correlation between color and variability amplitude remains an open question.

%Whether low-gravity and/or redder T dwarfs exhibit higher a variability remains unknown. Contrary to L dwarfs, near-infrared colors are not necessarily a good tracer of surface gravity (e.g., \citealt{Artigau:2011ly}). A handful of low-gravity T dwarfs are currently confirmed (e.g., \citealt{Gagne:2015d, Naud:2014}) and only one of them, GU Psc\,b, displays a low-significance variability on a 6\,h timescale \citep{Naud:2017a}.

%Furthermore, this excess of high-amplitude variables among redder dwarfs does not seem to hold at the L/T transition and among T dwarfs. Contrary to L dwarfs, in the T dwarf regime, $J-K$ colors are poor tracers of surface gravity , and the impact of surface gravity on the variability of cooler BDs will await the discovery, or identification as such among currently known objects, of a larger sample of low-gravity T dwarfs.

%}

\section{Spectroscopic Variability of Brown Dwarfs\label{spectro_var}}
While time-resolved photometry provides  constraints on the presence of weather-like patterns on brown dwarfs, time-resolved spectroscopy provides information on the physical nature of the processes at play. From the ground, {spectrophotometry} with sub-\% accuracy is very challenging, with variable slit-losses, telluric absorption and instrument flexures all masking low-level variability. The Wide-Field Camera 3 (WFC3; \citealt{MacKenty:2010}) slitless grism mode aboard the {\it Hubble Space Telescope} (HST) avoids these issues and provided the most compelling spectroscopic detections of variability in the literature. This mode covers the  1.1-1.7\,$\mu$m domain at a $\lambda/\Delta\lambda\sim100$ {resolving power}. It covers methane absorption longward of 1.6\,$\mu$m as well as a deep  water feature  centred at 1.4\,$\mu$m, a wavelength range that is largely inaccessible from the ground due to the Earth's atmospheric absorption. This WFC3 mode is also  central to exoplanet study, either through phase curve \citep{Stevenson:2014} or transit spectroscopy.

\citet{Buenzli:2012} reports the  detection of variability in the T6.5 2MASS\,J22282889--4310262 (2M2228) in WFC3 spectroscopy obtained simultaneously with Spitzer $4.5$\,$\mu$m photometry. 2M2228 was a known photometric variable with a relatively rapid $\sim$1.4\,h rotation period \citep{Clarke:2008, Radigan:2014}. The resulting light-curves  display sinusoidal modulation at various wavelengths with significant relative phase lags, by up to 180$^\circ$. These phase lags correlate with the effective pressure probed by each wavelength range (See Figure~3 in \citealt{Buenzli:2012}). A single spot on the surface of the BD would lead to a photometric modulation with a common phase at all wavelengths. A flux reversal, for example a  redder spot on a bluer  surface, can lead to an anti-correlation between wavelengths or a phase lag of 180$^\circ$. The phase lag at values other than 0$^\circ$or 180$^\circ$ would imply that the weather patterns in cause span a significant fraction of the circumference of the object. The typical scale height of a BD is on the order of a few km, while the radius is about  $7\times10^4$\,km. It would therefore seem unphysical for a single atmospheric feature than spans a few scale heights to be stretched half across the disk of the BD while preserving its  integrity. The authors suggest the presence of large-scale temperature and/or opacity gradients across the surface as a plausible explanation, but much detailed dynamical simulations will be required to draw any firm conclusion.

%\footnote{The scale height for an atmosphere is $kT/\mu g$. For a typical T dwarf with $\log g=5$ and T$_{\rm eff}$=900, this corresponds to 3.7\,km in an hydrogen-dominated atmosphere ($\mu=2$)}

%The dataset probes about an order of magnitude in pressure over its wavelength domain, and various cloud species, most notably sulfides, are expected to form at pressures and temperatures in that interval. 

\citet{Yang:2016a} presented a second set of observations of this mid-T, with simultaneous HST and Spitzer observations obtained two years later. Their observations show phase shifts similar to those reported by \citet{Buenzli:2012}, except for the 4.5\,$\mu$m bandpass. Light-curves in that dataset show phase lags clustering either around 0$^\circ$ or $180^\circ$ (See Figure~14 and 19 in \citealt{Yang:2016a}). This may not require significant extent of surface features and may be explained by a flux reversal within a single spot. This difference between two visits of the same objects highlights the fact that a better understanding of the range of behaviors seen on a single BD is needed before drawing firm conclusion regarding the differences between objects.

\begin{figure}
\includegraphics[width=0.5\paperwidth]{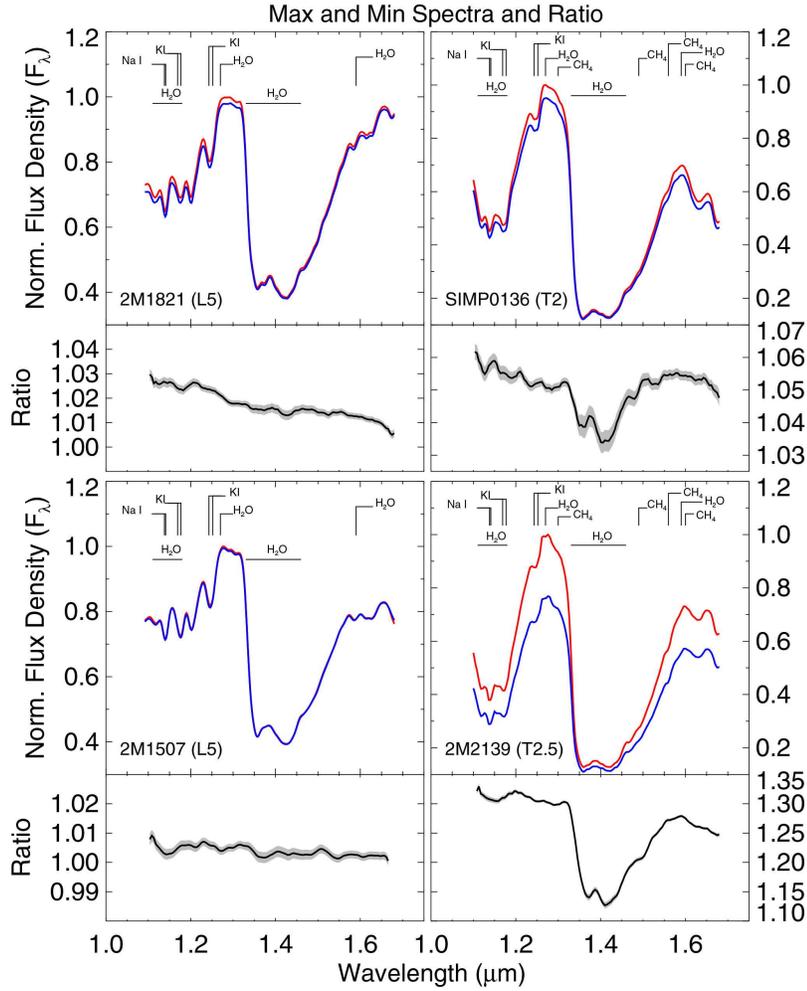}
\caption{   {\it [From \citet{Yang:2015}. Reproduced by permission of the AAS.]}  Spectral variability for two mid-L dwarfs (2M1821, 2M1507) and two early-T dwarfs (SIMP0136, 2M2139). The early-Ts, on the right, display a  lower variability in their water bands compared to other wavelengths, suggesting that the cloud decks involved in the photometric modulation lay at a depth intermediate between the altitude at which the $J$ and $H$-band flux is emitted, and that of water absorption. This behavior is not seen in L dwarfs, which is indicative of high-altitude clouds, above the depth of water absorption ($\sim$4 bars). The upper plot show the spectra corresponding to the maximum and minimum total fluxes of all four objects (respectively red and blue curves). The lower plots show the ratio of the maximum and minimum spectra; values close to unity indicate no variability.  }
\label{yang2015}    % Give a unique label
\end{figure}

\begin{figure}
\includegraphics[width=0.4\paperwidth]{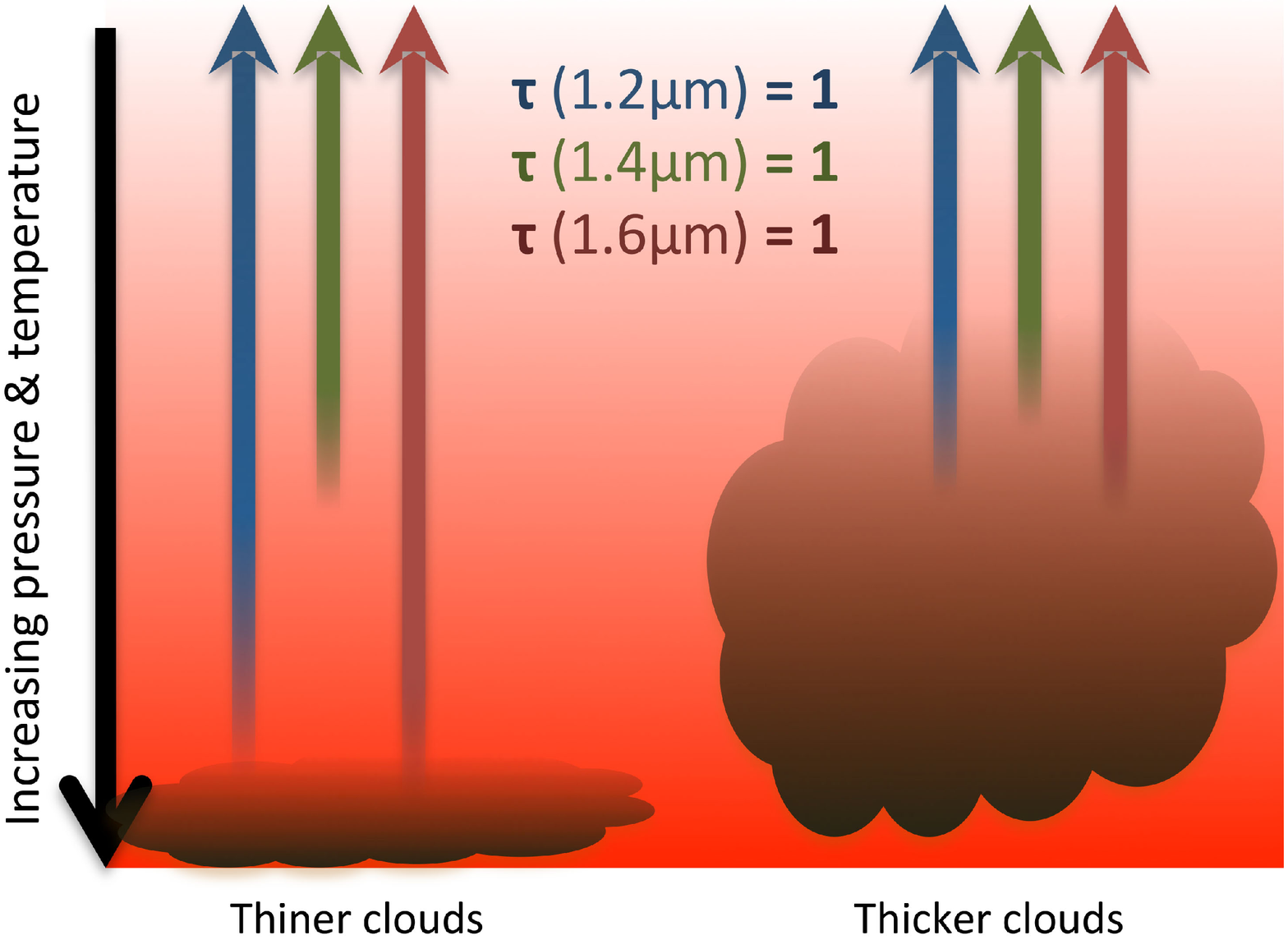}
\caption{Schematic representation of the cloud structure suggested by the results of \citet{Apai:2013} and \citet{Yang:2016a} .  In early T dwarfs, clouds of varying thicknesses are present on the surface and the  fractional coverage of thin and thick clouds changes through time. With thinner, low-lying cloud decks, the flux at 1.2\,$\mu$m and 1.6\,$\mu$m (i.e., the peak of $J$ and $H$ bands) probes deeper into the atmosphere while water bands at 1.4\,$\mu$m probes the cooler layers of the atmosphere. In the presence of thicker clouds, fluxes at  1.2\,$\mu$m and 1.6\,$\mu$m sample cooler, high-altitude cloud layers while the depth probed at 1.4\,$\mu$m does not change significantly. This leads to a higher variability at 1.2 and 1.6\,$\mu$m and lower variability at 1.4\,$\mu$m. }
\label{nuages}    % Give a unique label. 
\end{figure}

\citet{Apai:2013} present a dataset similar to the HST observations of \citet{Buenzli:2012} for two of the highest amplitude variable T dwarfs: SIMP0136 and 2M2139. Their variability was detected at a high significance, but no significant phase lag between spectral features probing different pressures in the atmosphere was observed. As shown in Figure~\ref{yang2015}, both objects show a decreased variability in the 1.4\,$\mu$m water absorption feature. This feature typically probes pressures of $\sim3$\,bar, while the middle of $J$ band probes pressures of $\sim10$\,bar \citep{Yang:2015}. This lower variability in the deep water bands implies that cloud decks that lead to photometric modulation in these two objects predominantly lay between these two pressures (See Figure~\ref{nuages}). This behavior is also seen in the slightly warmer T0.5 dwarf Luhman\,16B \citep{Buenzli:2015}. 

This consistent behavior between three early-Ts differs from the two L5 described in \citet{Yang:2015},  2MASS\,J$18212815+1414010$ and 2MASS\,J$15074769-1627386$ (2M1821, 2M1507), and the very dusty L6 dwarf WISE0047 \citep{Lew:2016}, where the variability within the water bands is similar to that of the $J$- and $H$-band peaks (see Figure~\ref{yang2015}).  This indicates that variability is due to hazes occurring above an optical depth of $\tau=1$ for water absorption in mid-L dwarfs. Interestingly, the effective pressure at the peak of $J$ and in the 1.4\,$\mu$m water feature  differ less for mid-L dwarfs (4.3 versus 6.5\,bar) than they do in early-T dwarfs (4.1 versus 8.1\,bar; see Table~6 in \citealt{Apai:2013}), which leads to a more wavelength-independent variability in {L dwarfs}. 

A noteworthy characteristic of the variability spectrum of 2M1821 (Figure~\ref{yang2015}) and in WISE0047 (See Figure~4 in \citealt{Lew:2016}) is the slope in the variability spectrum. Variability at $\sim$1.1\,$\mu$m is larger than at $\sim$1.6\,$\mu$m, with a linear trend in between. This behavior is best explained by a wavelength-dependent extinction within the high-altitude clouds and the slope provides information on the typical grain size within the clouds ($\sim0.4$\,$\mu$m; \citealt{Lew:2016}).

\section{Photometric Variability of Y Dwarfs}
The WISE mission \citep{Wright:2010} is an all-sky survey satellite that operated between $3.4$ and $22$\,$\mu$m and allowed for the first time to identify a sample of BDs well below 600\,K, corresponding to the Y spectral class \citep{Cushing:2011,Kirkpatrick:2012}.  At such low temperatures, {Y dwarfs} are not expected to host, silicate-bearing dust grains close to or above their photosphere such as what is seen in L and early T dwarfs. A number of chemical species are nevertheless expected to form clouds in cool atmosphere, such as sulfides \citep{Morley:2012}, or for the coolest objects, ammonia and water \citep{Skemer:2016a, Leggett:2015}. The coolest of these objects, such as  WISE J085510.83-071442.5 (W0855; \citealt{Luhman:2016, Luhman:2014}), may well host weather patterns that include the rain and snow that are familiar to earthlings, albeit in a completely different physical setting.

From the ground, the faintness of {Y dwarfs} in the near-infrared and the overwhelming thermal background in the mid-infrared strongly limit the possibilities to study their variability. Only Spitzer is sufficiently sensitive beyond 3\,$\mu$m to allow {Y dwarf} studies at high photometric accuracy. \citet{Cushing:2016} and \citet{Leggett:2016a} reported preliminary results of a {Y dwarf} variability survey, with the detection of similar variability amplitudes ($\sim$3\% at 4.5\,$\mu$m) and periods (6$-$8.5\,h) in WISEP\,J$140518.40+553421.4 (W1405)$ and WISEP\,J$173835.53+273258.9$. The very red 3.6\,$\mu$m to 4.5\,$\mu$m color of {Y dwarfs} makes variability detection at 3.6\,$\mu$m challenging, and only one epoch of the W1405 observations shows an unambiguous detection at 3.6\,$\mu$m, leading to a 3.6 to 4.5\,$\mu$m amplitude ratio close to unity. The variability of these {Y dwarfs}, both in terms of amplitude and timescale, is comparable to that of {T dwarf} as measured by \citet{Metchev:2015}. However, the important difference in temperature suggests that different cloud species are most likely at play.
%  WISEP J173835.52+273258.9    ->>>> Leggett 2016
% WISE J140518.39+553421.3  -->>> Cushing:2016

\begin{figure}
\includegraphics[width=\textwidth]{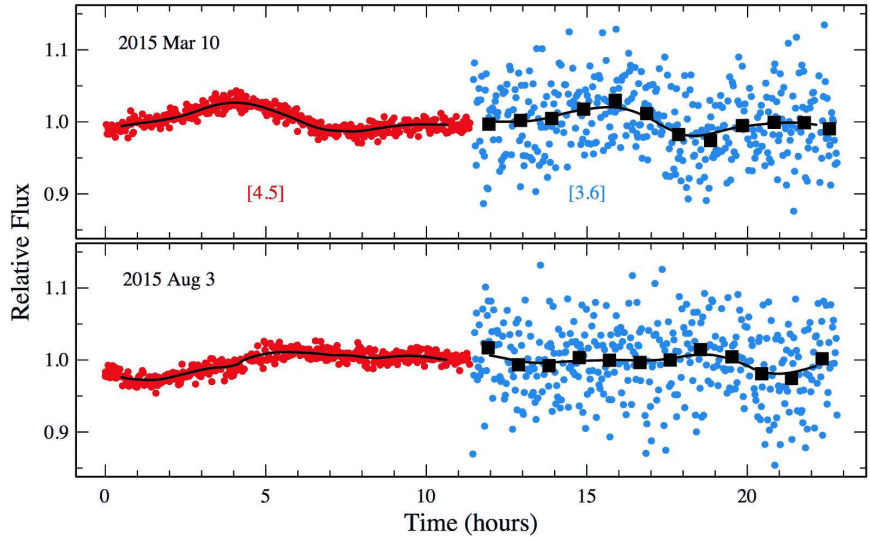}
\caption{ {\it [From \citet{Esplin:2016}. Reproduced by permission of the AAS.]} Spitzer IRAC timeseries of W0855 at 4.5\,$\mu$m and 3.6\,$\mu$m (respectively red and blue dots). The 3.6\,$\mu$m filter probes a deep methane absorption band and W0855 is much fainter at this wavelength than at 4.5\,$\mu$m, hence the lower photometric precision.}
\label{esplin2016}    % Give a unique label
\end{figure}

The Y dwarf W0855 has a temperature of only $250$\,K and an estimated mass below the deuterium burning limit \citep{Luhman:2014}. It is a free-floating analog to the cold evolved giant planets found by radial-velocity (RV) surveys and provides a unique opportunity to understand their atmospheres. This brown dwarf was monitored by \citet{Esplin:2016} on two epochs with Spitzer and clear photometric variability was detected at $3.6$\,$\mu$m and $4.5$\,$\mu$m (See Figure~\ref{esplin2016}). While the variability is unambiguous, no accurate period can be measured because the evolution of the light-curve masks any clear periodicity. The best estimates suggest a $9-14$\,h rotation, comparable to Solar System gas giants. The variability amplitude ratio between the two bandpasses is  close to unity, similar to the two warmer Y dwarfs mentioned above. While it could be tempting to attribute this variability to water ice clouds, the data in hand for W0855 falls short from confirming this hypothesis and only time-resolved spectroscopy could establish whether we are witnessing our first BD snowstorm. 

With a collection area 50 times larger and a much more diverse suite of observing modes, the James Webb Space Telescope (JWST) will provide vastly improved constraints on the nature of Y dwarf variability compared to what is currently possible with Spitzer data. The predicted sensitivities of JWST's Near Infrared Spectrograph (NIRSpec) should allow spectroscopy at a resolving power of $\sim1000$ and a signal-to-noise ratio above 100 per resolution element at $4$\,$\mu$m for an hour-long integration on W0855. This will lead to a detection of spectroscopic variability within each resolution element, assuming a variability level comparable to that reported by \citet{Esplin:2016} and will allow to perform detailed modelling of cloud dynamics and chemistry.
%  https://jwst.stsci.edu/files/live/sites/jwst/files/home/science%20planning/performance%20and%20simulation%20tools/sensitivity%20overview/_documents/R1000_sens.txt
%  R=1000 Gratings
%  Limiting point source sensitivity
%  100.000 s total exposure
%  1.000 s sub-exposures
%  S/N=10 per 2 pixel wide spectral resolution element
%
%  Wave   Spect.  Cont.  Line
%  length  res.   Flux   Flux
%  (micron)  R    (nJy) (erg/s/cm2)
%
%  4.00  1012.7   336.4  2.49E-19
% [4.5] = 13.89
% sqrt(2.431e-12*10^(-13.89/2.5)/(2.49d-19))*10.0*sqrt(60/10.) = 127 ... SNR à R=1000 pour une heure sur WISE0855

\section{Brown Dwarf Variability as a Limitation to Radial-Velocity Surveys}
Brown dwarf variability provides a wealth of information on weather patterns that would be difficult or impossible to obtain otherwise, but these patterns will also pose significant challenges to other aspects of BD study. BDs are hosts to planetary-mass companions, which are often referred to as planets even if the formal IAU definition states that ``planets'' orbit stars. The first directly imaged planet was found around the young BD 2MASSW\,J$1207334-393254$  \citep{Chauvin:2005a}. Disks are also common around young BDs (e.g., \citealt{Luhman:2005}), suggesting that planet formation is commonplace around these objects.  BDs are enticing targets for radial-velocity (RV) searches as planets will induce a much larger signal than for Sun-like stars due to the lower mass of the host ($<7$\,\% of the Sun). Furthermore, the small radii of BDs would ease the detection and characterization of eventual transiting planets \citep{Belu:2013}. From signal-to-noise considerations alone, the upcoming and recently-commissioned precision RV spectrographs (e.g., CARMENES, IRD, NIRPS or SPIRou; \citealt{Quirrenbach:2014, Tamura:2012, Bouchy:2017, Artigau:2014d}) will have the sensitivity to search for terrestrial planets around BDs, but their variability is likely to be a significant limitation. \citet{Metchev:2015} found that about half of {L dwarfs} display $>0.2\%$ variability in the {Spitzer} bandpasses. With a majority of {L dwarfs} rotating faster than 20\,km/s, and typically having at least $0.2$\,\% variability, one should expect a typical variability-induced jitter of $>40$\,m/s. This  is analogous to the activity jitter encountered with M dwarfs, a problem that has received significant attention as RV surveys extend to ever cooler targets \citep{Boisse:2011, Reiners:2010a}. While hampering future RV planet searches around L and T dwarfs, weather-induced RV jitter opens the door to  {Doppler imaging} \citep{Vogt:1983}  as a new technique for exploring the atmosphere of BDs.

\section{Doppler Imaging of Brown Dwarfs}
\begin{figure}
\includegraphics[width=0.5\paperwidth]{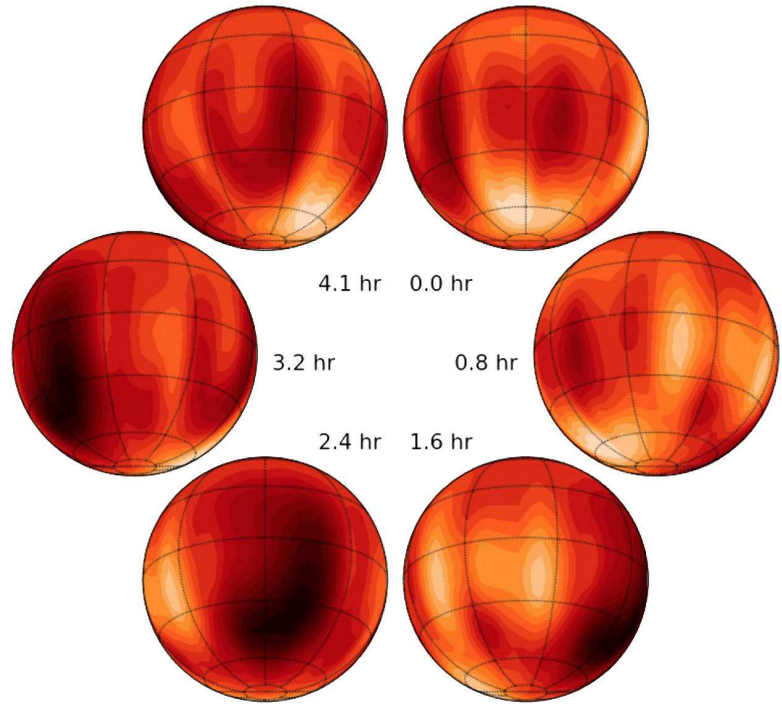}

\caption{  {\it [Reprinted by permission from Springer Nature: A global map of the nearest known brown dwarf, \citealt{Crossfield:2014}.]} Global surface brightness map of Luhman\,16B derived from Doppler imaging. Each of the 6 maps shows the observer-facing hemisphere through the 4.9\,h rotation period. A dark region close to the equator (2.4\,h) and a bright region close to the pole (0.0\,h) are recovered at high significance. The contrast between the darkest and brightest regions is $\pm10$\%.} 
\label{crossfield2014}    % Give a unique label
\end{figure}

The main motivation behind the study of BD variability is to have a glance at the diversity of weather patterns on their surfaces. Doppler imaging is a well established technique that has been used to resolved stellar features for decades. As a star rotates, brightness variations on its surface translate into time-varying signatures in its mean spectral line profile. This technique has been extended to active M dwarfs \citep{Barnes:2001} and BDs are promising targets for Doppler imaging studies. The presence of cloud patterns is well established through photometric variability and their rotation profiles can be resolved by state-of-the-art RV spectrographs operating in the near-infrared. The rich molecular bands are amenable to least-square deconvolution, which partially offsets the loss in signal-to-noise due to their relative faintness. \citet{Crossfield:2014} presented the first demonstration of Doppler imaging on the T0.5 dwarf Luhman\,16B. This objects is one of the best possible cases for this type of study as it is much brighter than any other T dwarf due to its proximity (2.0\,pc; \citealt{Luhman:2013}) and displays one of the largest known photometric variabilities among T dwarfs \citep{Buenzli:2015,Biller:2013c}. The map was reconstructed using only a relatively short wavelength interval centered on the 2.29\,$\mu$m CO bandhead, with CRIRES at the VLT \citep{Kaeufl:2004}. The recovered map (see Figure~\ref{crossfield2014}) shows large-scale inhomogeneities in surface brightness; whether these represent differences in temperature or composition remains to be seen. These results represent an exciting proof-of-concept as a tool to probe BD atmosphere. Obtaining simultaneous maps of various chemical species with strong near-infrared signatures (e.g., methane or water) is possible, as well as a multi-epoch monitoring of cloud maps. These will yield strong constraints on atmosphere dynamics of BDs. A few brighter M/L transition dwarfs and a handful {L dwarfs} will be amenable to Doppler imaging with $4-8$\,m class telescopes equipped with broad-band precision radial-velocity spectrographs in the near-infrared. The advent of similar instruments on 30\,m-class telescopes will pave the way to surface mapping of dozens of BDs and possibly the  brighter imaged exoplanets \citep{Crossfield:2014a}.

%Surface map of brown dwarf Luhman 16B, which clearly depicts a bright
%near-polar region (seen in the upper-right panels) and a darker mid-latitude area
%(lower-left panels) consistent with large-scale cloud inhomogeneities. The lightest
%and darkest regions shown correspond to brightness variations of roughly ±10%. The
%time index of each projection is indicated near the center of the figure.

% IF you do NOT use bibtex, put comments before the following 2 lines
%\bibliographystyle{spbasicHBexo} %for bibtex
\bibliography{bibdesk} %for bibtex-example

\end{document}